\documentclass[letter]{aa}  

\usepackage{graphicx}
\usepackage{txfonts}

\begin{document}

   \title{Heading Gaia to measure atmospheric dynamics in AGB stars}

   \author{A. Chiavassa
          \inst{1}
          \and
          B. Freytag
          \inst{2}
          \and
          M. Schultheis
          \inst{1}
          }

 \institute{Universit\'e C\^ote d'Azur, Observatoire de la C\^ote d'Azur, CNRS, Lagrange, CS 34229, Nice,  France \\
		\email{andrea.chiavassa@oca.eu}
               \and
         Department of Physics and Astronomy at Uppsala University, Regementsv\"agen 1, Box 516, SE-75120 Uppsala, Sweden\\
             }

   \date{...}

 
  \abstract
   {Asymptotic Giant Branch (AGB) stars are characterised by complex stellar surface dynamics that affect the measurements and amplify the uncertainties on stellar parameters. As a matter of fact, the uncertainties in observed absolute magnitudes originate mainly from uncertainties in the parallaxes. The resulting motion of the stellar photo-center could have adverse effects on the parallax determination with Gaia.}
   {We explore the impact of the convection-related surface structure in AGBs on the photocentric variability. We quantify these effects to characterise the observed parallax errors and estimate fundamental stellar parameters and dynamical properties.}
   {We use 3D radiative-hydrodynamics simulations of convection with
 CO5BOLD and the
   post-processing radiative transfer code {{\sc Optim3D}} to compute intensity
maps in the Gaia $G$ band [325 -- 1030~nm]. From those maps, we calculate the intensity-weighted mean of all emitting points tiling the visible stellar surface (i.e., the photo-center) and evaluate its motion as a function of time. We extract the parallax error from Gaia DR2 for a sample of semiregular variables in the solar neighbourhood and compare it to the synthetic predictions of photo-center displacements.}
   {AGB stars show a complex surface morphology characterised by the presence of few large scale long-lived convective cells accompanied by short-lived and small scale structures. As a consequence, the position of the photo-center displays temporal excursions between 0.077 to 0.198~AU ($\approx$5 to $\approx$11$\%$ of the corresponding stellar radius), depending on the simulation considered.  We show that the convection-related variability accounts for a substantial part to the Gaia DR2 parallax error of our sample of semiregular variables. Finally, we put in evidence for a correlation between the mean photo-center displacement and the stellar fundamental parameters: surface gravity and pulsation.  We denote that parallax variations could be exploited quantitatively using appropriate RHD simulations corresponding to the observed star.}
   {}

   \maketitle
%

    \begin{figure*}[!h]
   \centering
    \begin{tabular}{c}
      \includegraphics[width=0.45\hsize]{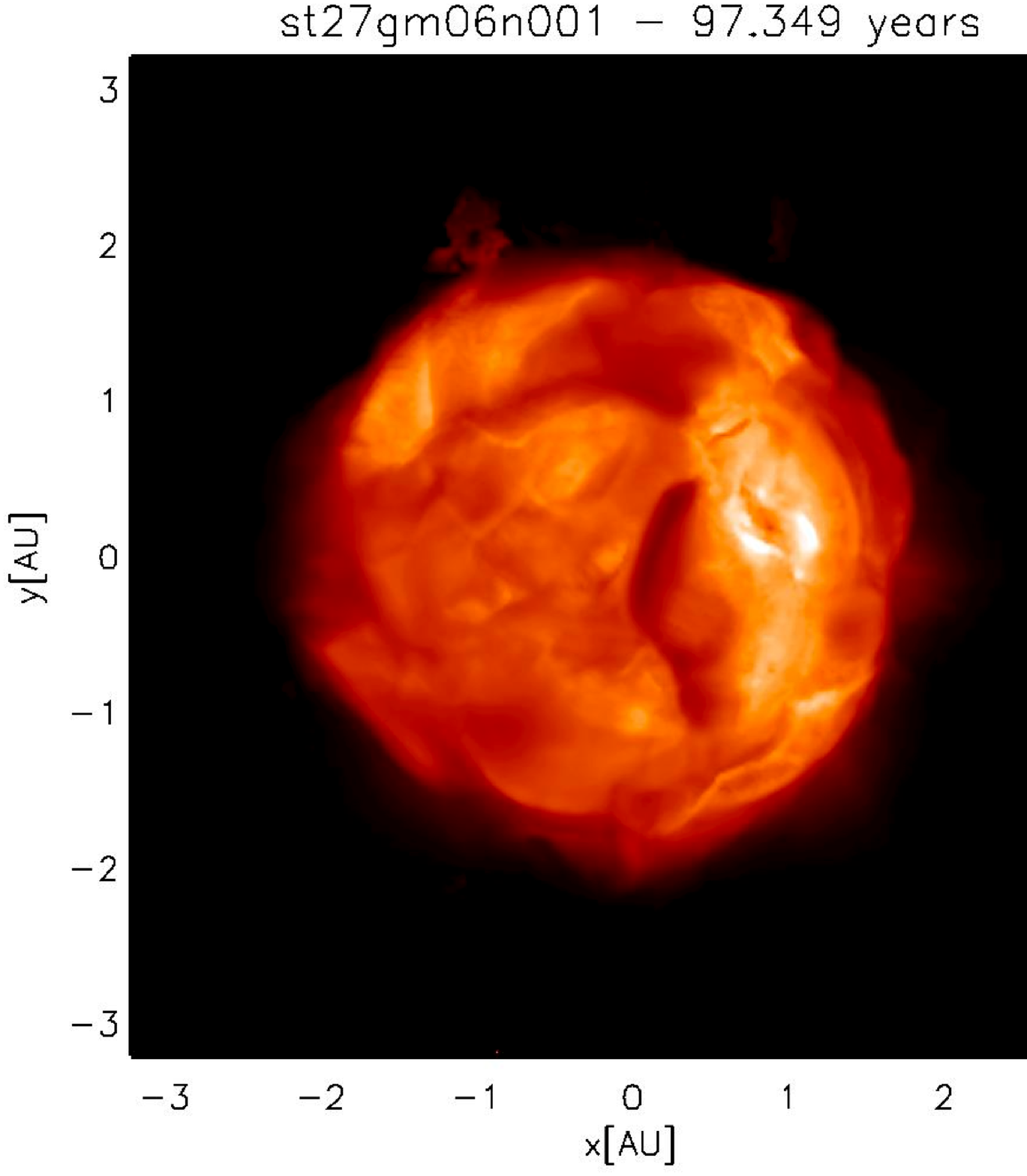}  
      \hspace{1.cm}
      \includegraphics[width=0.45\hsize]{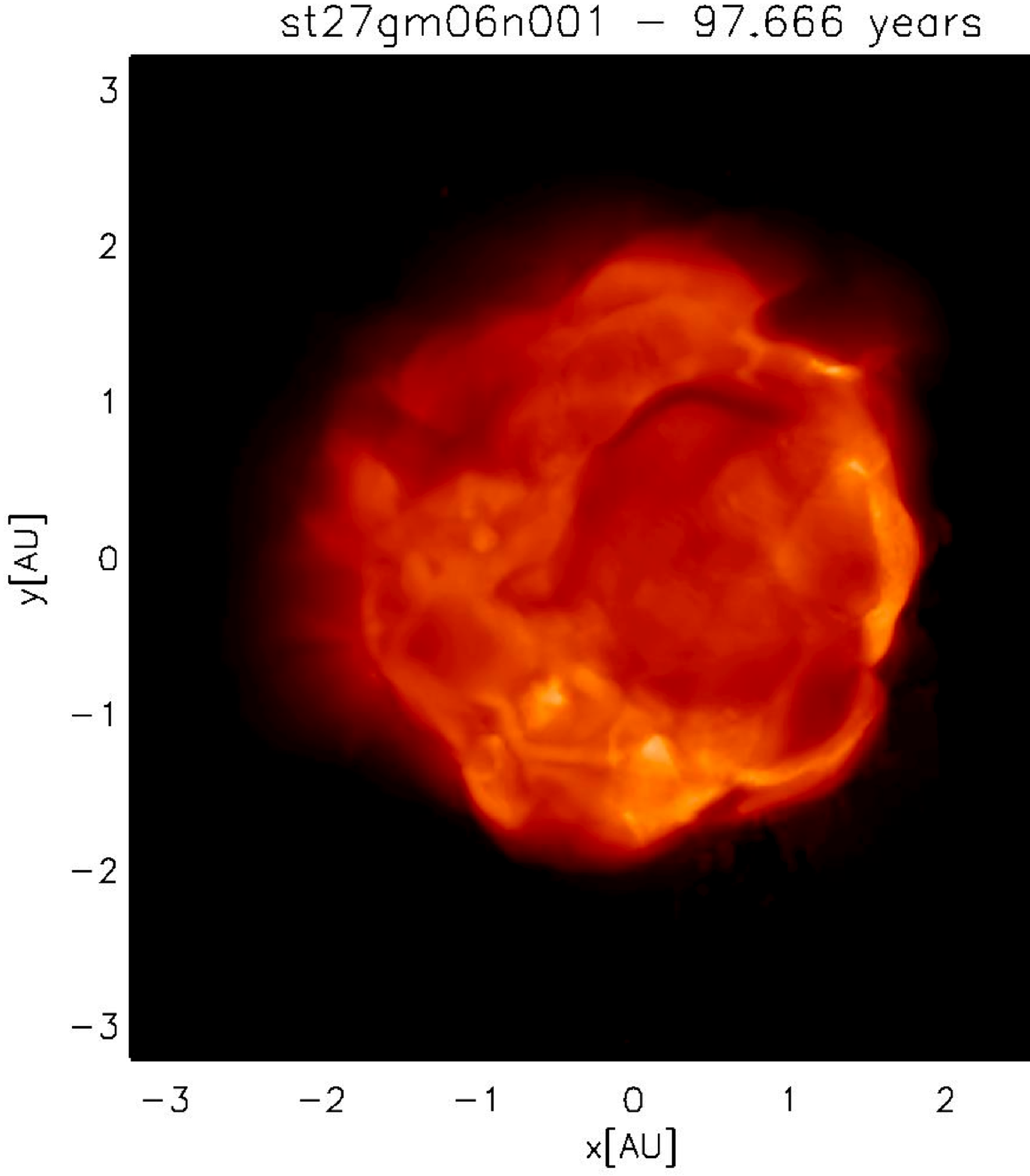}  
  \end{tabular}
      \caption{Example of the squared root intensity maps (the range is $\sqrt{[0.-3000.]}$
        erg/s/cm$^2$/\AA ) in the Gaia $G$ photometric system \citep{2018arXiv180409368E} for two different snapshots of one simulation listed in Table~\ref{simulations}. The number on the top indicates the stellar times at which the two snapshots were computed.}
        \label{imAGBs}
           \end{figure*}   

\section{Introduction}
AGBs are low- to intermediate-mass stars that evolve to red giant and asymptotic giant branch increasing the mass-loss during this evolution. They are characterised: (i) by large amplitude variations in radius, brightness and temperature of the star; (ii) and by a strong mass loss rate driven by an interplay between pulsation, dust formation in the extended atmosphere, and radiation pressure on the dust \citep{2018A&ARv..26....1H}. Their complex dynamics affect the measurements and amplify the uncertainties on stellar parameters.

Gaia \citep{2016A&A...595A...1G} is an astrometric, photometric and spectroscopic space borne mission. It performs a survey of a large part of the Milky Way. The second data release (Gaia DR2) in April 2018 \citep{2018arXiv180409365G} brought high-precision astrometric parameters (i.e., positions, parallaxes, and proper motions) for over 1 billion sources brighter that $G\approx20$. Among all the objects that have been observed, the complicated atmospheric dynamics of AGB stars affect the photocentric position and, in turn, their parallaxes \citep{2011A&A...528A.120C}. The convection-related variability, in the context of Gaia astrometric measurement, can be considered as "noise" that must be quantified in order to better characterise any resulting error on the parallax determination. However, important information about stellar properties, such as the fundamental stellar parameters, may be hidden behind the Gaia measurement uncertainty. 

In this work we explore the effect of convection-related surface structures on the photo-center to estimate its impact on the Gaia astrometric measurements.

\section{Methods}

We used the radiation-hydrodynamics (RHD) simulations of AGB stars \citep{2017A&A...600A.137F} computed with CO$^5$BOLD \citep{2012JCoPh.231..919F} code. The code solves the coupled non-linear equations of compressible hydrodynamics and non-local radiative energy transfer in the presence of a fixed external spherically symmetric gravitational field in  a 3D cartesian grid. It is assumed  that solar abundances are appropriate for M-type AGB stars. The basic stellar parameters of the RHD simulations are reported in Table~\ref{simulations}. The configuration used is the "star-in-a-box", where the evolution outer convective envelope and the inner atmosphere of AGB stars are taken into account in the calculation. In the simulations convection, waves, and shocks all contribute to the dynamical pressure and, thus, to an increase of the stellar radius and to a levitation of material into layers where dust can form. No dust is included in any of the current models. The regularity of the pulsations decreases with decreasing gravity as the relative size of convection cells increases. The pulsation period is extracted with a fit of the Gaussian distribution in the power spectra of the simulations. The period of the dominant mode increase with the radius of the simulation \citep[Table~1 in ][]{2017A&A...600A.137F}.
     
\begin{table*}[htb]
\footnotesize 
\begin{center}
 \caption{RHD simulation parameters
 \label{simulations}}
 \begin{tabular}{l|rrrrrrrr|rrrr}
\hline
Simulation & $M_\star$ & $L_\star$ & $R_\star$ & $T_\mathrm{eff}$ & $\log g$ & $t_\mathrm{avg}$ & $P_\mathrm{puls}$ & $\sigma_\mathrm{puls}$ & $\langle P\rangle$ & $\sigma_P$ & $\langle P_x\rangle$ & $\langle P_y\rangle$ \\
& $M_\sun$ & $L_\sun$  & AU & K & (cgs) & yr & yr & yr &  AU & AU  & AU & AU \\ \hline
st26gm07n002 & 1.0 &   6986 &  2.04 & 2524 & -0.85 &  25.35  & 1.625 & 0.307 &  0.262 & 0.187 & -0.100 &   0.046 \\
st26gm07n001 & 1.0 &  6953 &   1.87 & 2635 & -0.77 & 27.74 & 1.416 & 0.256 &  0.275 & 0.198 & -0.098  &  0.024\\
st28gm06n26 & 1.0 &   6955 &    1.73 & 2737 & -0.70 & 25.35 & 1.290 & 0.317 & 0.241 & 0.152 & -0.068 & -0.002 \\
st29gm06n001 & 1.0 &  6948 &    1.62 & 2822 & -0.65 & 25.35 & 1.150 & 0.314 & 0.266 & 0.174 & -0.098  & 0.016 \\
st27gm06n001 & 1.0 &   4982 &    1.61 & 2610 & -0.64 & 28.53 & 1.230 & 0.088 & 0.150 & 0.101 & -0.027 & 0.027 \\ 
st28gm05n002 & 1.0 & 4978 &    1.46 & 2742 & -0.56 & 25.35 & 1.077 & 0.104 & 0.133 & 0.077  & -0.002 &  0.033\\
st28gm05n001 & 1.0 &   4990 &    1.40 & 2798 & -0.52 & 25.36 & 1.026 & 0.135 & 0.183 & 0.131 & -0.057  & 0.0174 \\
st29gm04n001 & 1.0 &   4982 &    1.37 & 2827 & -0.50 & 25.35 & 0.927 & 0.100 & 0.152 & 0.078 & -0.002 & 0.023 \\
\hline
 \end{tabular}
\end{center}
{The table shows the simulation name, the mass $M_\star$, the average emitted luminosity $L_\star$, the average approximate stellar radius $R_\star$ (note that the radii vary by about 20$\%$ during one pulsation period), effective temperature $T_\mathrm{eff}$, and surface gravity $\log g$, the pulsation period $P_\mathrm{puls}$, the half of the distribution of the pulsation frequencies $\sigma_{\mathrm{puls}}$, the stellar time $t_\mathrm{avg}$ used for the averaging of the rest of the quantities. All these quantities are from \cite{2017A&A...600A.137F}. The last four columns are the time-averaged value of the photo-center displacement $\langle P\rangle=\langle (P_x^2 + P_y^2)^{(1/2)}\rangle$, its standard deviation ($\sigma_P$), and the time-averaged of $P_x$ and $P_y$.}
\end{table*}

We computed intensity maps based on snapshots from the RHD simulations integrating in the Gaia $G$ photometric system \citep{2018arXiv180409368E}. For this purpose, we employed the code
{{\sc Optim3D}} \citep{2009A&A...506.1351C}, which takes into account the Doppler shifts caused by the
convective motions. The radiative transfer is computed in detail using pre-tabulated extinction coefficients per unit mass, as for MARCS (\citealp{2008A&A...486..951G}) as a function of temperature, density, and wavelength for the solar composition \citep{2009ARA&A..47..481A}. Micro-turbulence broadening was switched off. The
temperature and density distributions are optimised to cover the values
encountered in the outer layers of the RHD simulations. \\
The surface of the deep convection zone has large and small convective cells. The visible fluffy stellar surface is made of shock waves, that are produced in the interior and that are shaped by the top of the convection zone as they travel outward \citep{2017A&A...600A.137F}. In addition to this, on the top of the convection-related surface structures, other structures appears. They result from the opacity effect and dynamics at Rosseland optical depths smaller than~1 (i.e., further up in the atmosphere with respect to the continuum-forming region). At the wavelengths in Gaia $G$-band, TiO molecules produce strong absorption. Both effects affect the position of the photo-center and causes temporal fluctuations during the Gaia mission, as already pointed out for red supergiant stars in \cite{2011A&A...528A.120C}. In the Gaia $G$ photometric system (Fig.~\ref{imAGBs}), the situation is analogous with few large convective cells with size of a third of the stellar radii (i.e., $\approx0.6$ AU) that evolve on a scale of several months to a few years and short-lived (weeks to month) and small scale ($\lesssim10\%$ of the stellar radius). The resulting position of the photo-center is affected with temporal fluctuations during the Gaia mission, as already pointed out for red supergiant stars in \cite{2011A&A...528A.120C}.\\
We calculated the position of the photo-center for each map (i.e., as a function of time) as the intensity-weighted mean of the $x-y$ positions of all emitting points tiling the visible stellar surface according to\\
\begin{eqnarray}
P_x=\frac{\sum_{i=1}^{N} \sum_{j=1}^{N} I(i,j)*x(i,j)}{\sum_{i=1}^{N} \sum_{j=1}^{N} I(i,j)} \\
P_y=\frac{\sum_{i=1}^{N} \sum_{j=1}^{N} I(i,j)*y(i,j)}{\sum_{i=1}^{N} \sum_{j=1}^{N} I(i,j)},
\end{eqnarray}
where $I\left(i,j\right)$ is the emerging intensity for the grid
point $(i,j)$ with coordinates $x(i,j)$, $y(i,j)$ of the
simulation, and $N=281$ is the total number of grid points in the simulated box. In presence of surface brightness asymmetries, the photo-center position will not coincide with the barycenter of the star and its position will change as the surface pattern changes with time. This is displayed in the photo-center excursion plots for each simulation in the Appendix together with the time-averaged photocenter position ($\langle P\rangle$) and its standard deviations ($\sigma_P$) in Table~\ref{simulations}. The value of $\sigma_P$ (the third-last one in Table~\ref{simulations}) is mostly fixed by the short time scales corresponding to the small atmospheric structures. However, the fact that $\langle P_x\rangle$ and $\langle P_y\rangle$ do not average to zero (last two columns of Table~\ref{simulations} and, e.g., Fig.~\ref{photo6} in the Appendix), means that the photo-center tends not to be centred most of the time on the nominal center of the star, because of the presence of a large convective cell. 

$\sigma_P$ varies between  0.077 to 0.198~AU ($\approx$5 to $\approx$11$\%$ of the corresponding stellar radius) depending on the simulation. This measure of the mean photo-center noise induced by the stellar dynamics in the simulations is compared in the next section to Gaia measurement uncertainty to extract information on stellar parameters from the astrometric measurements. It should be noted that the main information that is used to determine the
astrometric characteristics of every star will be the along-scan measurement of Gaia. \cite{2011A&A...528A.120C} showed that the projection of the star position along the scanning direction of the satellite with respect to a known reference point disclose similar, though slightly increased, values of $\sigma_P$. At the current state of the DR2, it is not possible to perform this on the data and we assumed the conservative value of $\sigma_P$ directly extracted from the RHD simulations for the following comparisons.

\section{Observations}

Evolved late-type stars show convection-related variability that may be considered, in the context of Gaia astrometric measurement, as "noise". \cite{2011A&A...528A.120C} demonstrated that RHD simulations can account for a substantial part of the supplementary 'cosmic noise', that affects Hipparcos measurements, for some prototypical red supergiant stars. As a consequence, the convection-related noise has to be quantified in order to better characterise any resulting error on the parallax.

We extracted the parallax error ($\sigma_\varpi$) from Gaia DR2 for a sample of semiregular variables (SRV) from \citet{tabur2009}, \citet{glass2007}, and \citet{1993ApJ...413..298J} that match the theoretical luminosities of RHD simulations (Table~\ref{simulations}). Moreover, we only considered stars with 4000 < $L_{\odot}$ < 8000 in order to compare with our simulations. It has to be noted that $\sigma_\varpi$ may still vary in the following data releases because: (i) the mean number of measurements for each source amounts to 26 \citep{2018arXiv180502035M} and it will be 70-80 times in total at the end of the nominal mission; (ii) and new solutions may be applied to adjust the imperfect chromaticity correction \citep{2018arXiv180409375A}.
 We cross-identified our sample stars with the Gaia DR2 release as well as with the distance catalog of \citet{Bailer-Jones2018} that uses a weak distance prior, varying as a function of Galactic longitude and latitude, to derive distances for the Gaia DR2 release\footnote{We used the TAP service at http://gaia.ari.uni-heidelberg.de/tap.html}. The apparent  $K$-band magnitudes were transformed to absolute $K$ magnitudes using the Gaia distances of \citet{Bailer-Jones2018} and neglecting the interstellar absorption which should be very small in the local neighbourhood. The absolute $K$ magnitudes were converted to bolometric magnitudes using the bolometric correction formula of \citet{Kerschbaum2010}. 
 The typical uncertainties in the bolometric correction ($\rm BC_{K}$)
are in the order of $\rm \pm 0.1\,mag$.  Finally, the  bolometric magnitudes where converted  to  luminosities assuming 
a solar $\rm Mbol_{\odot} =4.7$ (\citealt{torres2010}). The error bars on luminosities were calculated by using the upper and lower distance limits provided by \citet{Bailer-Jones2018} for each of our sources together with the error of $\rm \pm 0.1\,mag$ for $\rm BC_{K}$.  We do not correct for the variation of the $K$-band light curve as only amplitudes in the visual are
available and the $K$-band amplitude of AGB stars is in general much smaller than in the visual.


   \begin{figure*}[!h]
   \centering
    \begin{tabular}{c}
    \includegraphics[width=0.33\hsize]{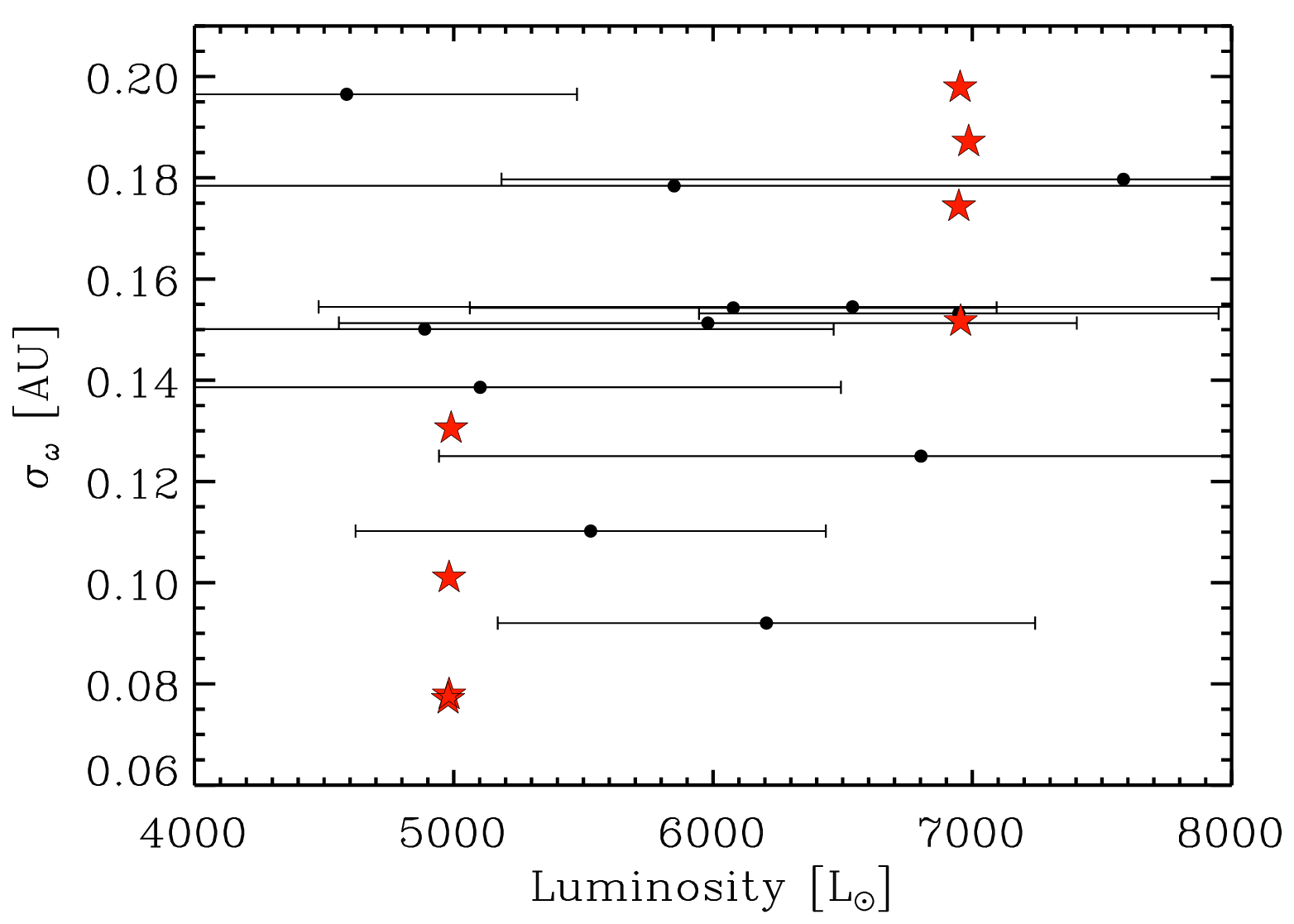} 
     \includegraphics[width=0.33\hsize]{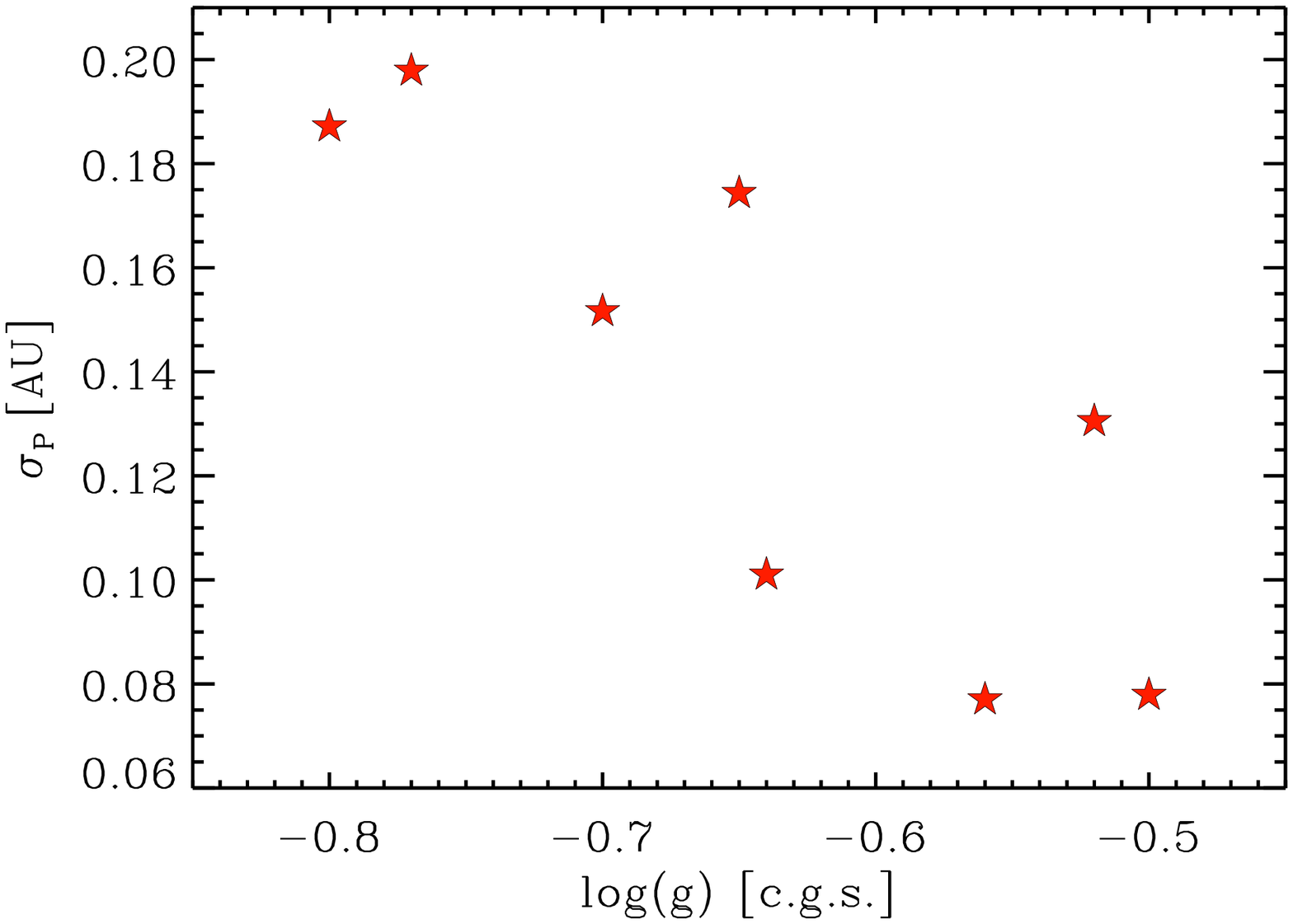} 
      \includegraphics[width=0.33\hsize]{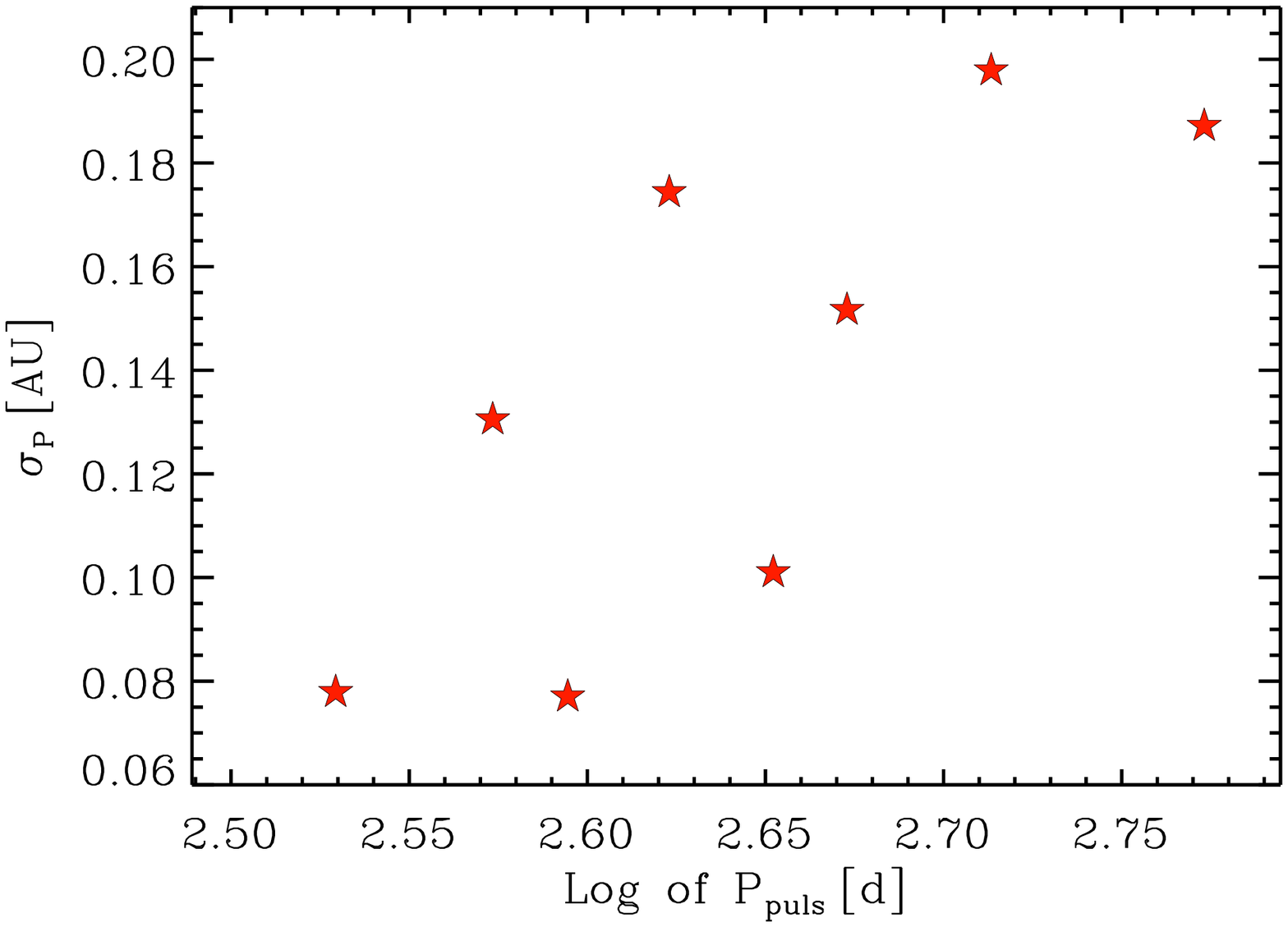} 
  \end{tabular}
      \caption{\emph{Left panel:} luminosity against the parallax error of the observations ($\sigma_\varpi$, circle symbol in black) and the standard deviation of the photo-center displacement for the RHD simulations of Table~\ref{simulations} ($\sigma_P$, star symbol in red). \emph{Central panel: }$\sigma_P$ against the surface gravity for the RHD simulations. \emph{Right panel: }$\sigma_P$ against the logarithm of the period.}
        \label{im1}
           \end{figure*}

\section{Comparison and predictions}

AGB stars are characterised by complex stellar surface dynamics that affects the measurements and on the determined stellar parameters. As a matter of facts, the uncertainties in observed absolute magnitudes originate mainly from uncertainties in the parallaxes. In this Section we investigate if the parallax errors of our SRV sample can be explained by the resulting motion of the stellar photo-center seen in the RHD simulations.

Fig.~\ref{im1} (left panel) displays the parallax errors against the luminosity and compares these results to the standard deviations of the photo-center displacement in the simulations. The latter show a good agreement with the observations. There are two luminosity families in the 3D simulations. In general the more luminous models are
larger, i.e., have lower surface gravity, which causes larger convection cells and a more fluffy atmosphere \citep{2017A&A...600A.137F}. As a consequence, for the simulation with higher luminosity ($\approx$7000 $L_\sun$), the parallax error ratio, defined as $\langle\sigma_\varpi\rangle$/$\sigma_P$ (where $\langle\sigma_\varpi\rangle$ is the average error for the stars with luminosity close enough to the corresponding simulation luminosity), is [0.75-0.98]. This attests that convection-related variability accounts for a substantial part to the parallax error in Gaia measurements. For lower luminosities ($\approx$5000 $L_\sun$), the ratio is [1.15-1.95] indicating that for those simulations the question is less clear. However, the observed and simulated luminosities do not coincide exactly and the observed error bars are still very large. One limitation of the existing model grid is the restriction to 1 $M_\sun$. In the future, there will be models with other masses available. For a more better comparison, we need simulations and  observations with known luminosities, masses, and radii. That is not easy on both the theoretical and the observational side.

An important piece of information is indeed hidden in the Gaia measurement uncertainty: using the corresponding stellar parameters of the RHD simulations of Table~\ref{simulations}, we plotted (Fig.~\ref{im1}) $\sigma_P$ as a function of surface gravity (central panel) and pulsation (right panel). They display a correlation between the mean photo-center displacement and the stellar fundamental parameters. While effective temperature does not show particularly correlated points, simulations with lower surface gravity (i.e., more extended atmospheres) return larger excursions of the photo-center. This behaviour is explained by the correlation between the stellar atmospheric pressure scale height ($H_{\rm{p}}\approx\frac{T_{\rm eff}}{g}$) and the photocenter displacement \citep{2001ASPC..223..785F,2006A&A...445..661L,2011A&A...528A.120C}.

An AGB star's property that is well constrained by observations is the Period-Luminosity (P-L) relation. The uncertainties in the determination of this relation is mainly based on the calculation for the distances and on the different P-L relations used. Fig.~\ref{im1} (right panel) reveals the correlation between the photo-center displacement and the logarithm of the pulsation: larger values of $\sigma_P$ correspond to longer pulsation periods. Global shocks induced by large-amplitude, radial and fundamental-mode pulsations have and impact on the detailed stellar structure of the stellar atmosphere, together with small-scale shocks. Both contribute to the levitation of material and the detected photo-center displacement \citep{2017A&A...600A.137F}. This result has to be associated with the P-L relation found by \cite{2017A&A...600A.137F}, who showed that the RHD simulations reproduce the correct period for a given luminosity compared to the observations of \cite{2009MNRAS.394..795W}.  Given the fact that $\sigma_P$ explains Gaia measurement uncertainties on the parallaxes (left panel), we denote that parallax variations from Gaia measurements could be exploited quantitatively using appropriate RHD simulations. However, the parameter space in our simulations is still limited (Table~\ref{simulations}). In the future we aim to extend our RHD simulations' parameters to lower and higher luminosities (i.e., shorter and longer periods) which will enable a more quantitative comparison with respect to the upcoming Gaia data releases.
    
\section{Summary and conclusions}

We used the snapshots from RHD simulations of AGB stars to compute intensity maps in the Gaia $G$ photometric system. The visible fluffy stellar surface is made of shock waves, that are produced in the interior and that are shaped by the top of the convection zone as they travel outward. The surface is characterised by the presence of few large and long-lived convective cells accompanied by short-lived and small scale structures. As a consequence, the position of the photo-center is affected by temporal fluctuations. 

We calculated the standard deviation of the photo-center excursion for each simulation and found that $\sigma_P$ varies between  0.077 to 0.198~AU ($\approx$5 to $\approx$11$\%$ of the corresponding stellar radius) depending on the simulation.
We compared the measure of the mean photo-center noise induced by the stellar dynamics in the simulations ($\sigma_P$) to the measurement uncertainty on the parallax of a sample of AGB stars in the solar neighbourhood cross-matched with the Gaia DR2 data. We found a good  agreement with observations probing that convection-related variability accounts for a substantial part to the parallax error. It has to be noted that $\sigma_\varpi$ may still vary in the following data releases because thanks to the increase of Gaia's measurements and further corrections to the parallax solution.

Finally, we put in evidence a correlation between the mean photo-center displacement and the stellar fundamental parameters: surface gravity and pulsation. Concerning the latter, we showed that that larger values of $\sigma_P$ correspond to longer pulsation periods. This result, associated with the P-L relation found by \cite{2017A&A...600A.137F} and the good agreement between simulations and observations ($\sigma_P$ versus $\sigma_\varpi$), let us denote that parallax variations from Gaia measurements could be exploited quantitatively using appropriate RHD simulations  corresponding to the observed star. 

\begin{acknowledgements}
This work has made use of data from the European Space Agency (ESA) mission Gaia (https://www.cosmos.esa.int/gaia), processed by the Gaia Data Processing and Analysis Consortium (DPAC, https://www.cosmos.esa.int/web/). Funding for the DPAC has been provided by national institutions, in particular the institutions participating in the Gaia Multilateral Agreement. AC et MS thank Patrick de Laverny for the enlightening discussions. The authors thank the referee for the helpful comments during the refereeing process.
\end{acknowledgements}

\bibliographystyle{aa}
\bibliography{biblio.bib}

\begin{thebibliography}{21}
\expandafter\ifx\csname natexlab\endcsname\relax\def\natexlab#1{#1}\fi

\bibitem[{{Arenou} {et~al.}(2018){Arenou}, {Luri}, {Babusiaux}, {Fabricius},
  {Helmi}, {Muraveva}, {Robin}, {Spoto}, {Vallenari}, {Antoja},
  {Cantat-Gaudin}, {Jordi}, {Leclerc}, {Reyl{\'e}}, {Romero-G{\'o}mez}, {Shih},
  {Soria}, {Barache}, {Bossini}, {Bragaglia}, {Breddels}, {Fabrizio},
  {Lambert}, {Marrese}, {Massari}, {Moitinho}, {Robichon}, {Ruiz-Dern},
  {Sordo}, {Veljanoski}, {Di Matteo}, {Eyer}, {Jasniewicz}, {Pancino},
  {Soubiran}, {Spagna}, {Tanga}, {Turon}, \& {Zurbach}}]{2018arXiv180409375A}
{Arenou}, F., {Luri}, X., {Babusiaux}, C., {et~al.} 2018, ArXiv e-prints

\bibitem[{{Asplund} {et~al.}(2009){Asplund}, {Grevesse}, {Sauval}, \&
  {Scott}}]{2009ARA&A..47..481A}
{Asplund}, M., {Grevesse}, N., {Sauval}, A.~J., \& {Scott}, P. 2009, \araa, 47,
  481

\bibitem[{{Bailer-Jones} {et~al.}(2018){Bailer-Jones}, {Rybizki}, {Fouesneau},
  {Mantelet}, \& {Andrae}}]{Bailer-Jones2018}
{Bailer-Jones}, C.~A.~L., {Rybizki}, J., {Fouesneau}, M., {Mantelet}, G., \&
  {Andrae}, R. 2018, ArXiv e-prints

\bibitem[{{Chiavassa} {et~al.}(2011){Chiavassa}, {Pasquato}, {Jorissen},
  {Sacuto}, {Babusiaux}, {Freytag}, {Ludwig}, {Cruzal{\`e}bes}, {Rabbia},
  {Spang}, \& {Chesneau}}]{2011A&A...528A.120C}
{Chiavassa}, A., {Pasquato}, E., {Jorissen}, A., {et~al.} 2011, \aap, 528, A120

\bibitem[{{Chiavassa} {et~al.}(2009){Chiavassa}, {Plez}, {Josselin}, \&
  {Freytag}}]{2009A&A...506.1351C}
{Chiavassa}, A., {Plez}, B., {Josselin}, E., \& {Freytag}, B. 2009, \aap, 506,
  1351

\bibitem[{{Evans} {et~al.}(2018){Evans}, {Riello}, {De Angeli}, {Carrasco},
  {Montegriffo}, {Fabricius}, {Jordi}, {Palaversa}, {Diener}, {Busso},
  {Cacciari}, \& {van Leeuwen}}]{2018arXiv180409368E}
{Evans}, D.~W., {Riello}, M., {De Angeli}, F., {et~al.} 2018, ArXiv e-prints

\bibitem[{{Freytag}(2001)}]{2001ASPC..223..785F}
{Freytag}, B. 2001, in 11th Cambridge Workshop on Cool Stars, Stellar Systems
  and the Sun, ed. {R.~J.~Garcia Lopez, R.~Rebolo, \& M.~R.~Zapaterio Osorio}
  (Astronomical Society of the Pacific Conference Series, Volume 223), 785

\bibitem[{{Freytag} {et~al.}(2017){Freytag}, {Liljegren}, \&
  {H{\"o}fner}}]{2017A&A...600A.137F}
{Freytag}, B., {Liljegren}, S., \& {H{\"o}fner}, S. 2017, \aap, 600, A137

\bibitem[{{Freytag} {et~al.}(2012){Freytag}, {Steffen}, {Ludwig},
  {Wedemeyer-B{\"o}hm}, {Schaffenberger}, \& {Steiner}}]{2012JCoPh.231..919F}
{Freytag}, B., {Steffen}, M., {Ludwig}, H.-G., {et~al.} 2012, Journal of
  Computational Physics, 231, 919

\bibitem[{{Gaia Collaboration} {et~al.}(2018){Gaia Collaboration}, {Brown},
  {Vallenari}, {Prusti}, {de Bruijne}, {Babusiaux}, \&
  {Bailer-Jones}}]{2018arXiv180409365G}
{Gaia Collaboration}, {Brown}, A.~G.~A., {Vallenari}, A., {et~al.} 2018, ArXiv
  e-prints

\bibitem[{{Gaia Collaboration} {et~al.}(2016){Gaia Collaboration}, {Prusti},
  {de Bruijne}, {Brown}, {Vallenari}, {Babusiaux}, {Bailer-Jones}, {Bastian},
  {Biermann}, {Evans}, \& et~al.}]{2016A&A...595A...1G}
{Gaia Collaboration}, {Prusti}, T., {de Bruijne}, J.~H.~J., {et~al.} 2016,
  \aap, 595, A1

\bibitem[{{Glass} \& {van Leeuwen}(2007)}]{glass2007}
{Glass}, I.~S. \& {van Leeuwen}, F. 2007, \mnras, 378, 1543

\bibitem[{{Gustafsson} {et~al.}(2008){Gustafsson}, {Edvardsson}, {Eriksson},
  {J{\o}rgensen}, {Nordlund}, \& {Plez}}]{2008A&A...486..951G}
{Gustafsson}, B., {Edvardsson}, B., {Eriksson}, K., {et~al.} 2008, \aap, 486,
  951

\bibitem[{{H{\"o}fner} \& {Olofsson}(2018)}]{2018A&ARv..26....1H}
{H{\"o}fner}, S. \& {Olofsson}, H. 2018, \aapr, 26, 1

\bibitem[{{Jura} {et~al.}(1993){Jura}, {Yamamoto}, \&
  {Kleinmann}}]{1993ApJ...413..298J}
{Jura}, M., {Yamamoto}, A., \& {Kleinmann}, S.~G. 1993, \apj, 413, 298

\bibitem[{{Kerschbaum} {et~al.}(2010){Kerschbaum}, {Lebzelter}, \&
  {Mekul}}]{Kerschbaum2010}
{Kerschbaum}, F., {Lebzelter}, T., \& {Mekul}, L. 2010, \aap, 524, A87

\bibitem[{{Ludwig}(2006)}]{2006A&A...445..661L}
{Ludwig}, H.-G. 2006, \aap, 445, 661

\bibitem[{{Mowlavi} {et~al.}(2018){Mowlavi}, {Lecoeur-Ta{\"i}bi}, {Lebzelter},
  {Rimoldini}, {Lorenz}, {Audard}, {De Ridder}, {Eyer}, {Guy}, {Holl},
  {Jevardat de Fombelle}, {Marchal}, {Nienartowicz}, {Regibo}, {Roelens}, \&
  {Sarro}}]{2018arXiv180502035M}
{Mowlavi}, N., {Lecoeur-Ta{\"i}bi}, I., {Lebzelter}, T., {et~al.} 2018, ArXiv
  e-prints

\bibitem[{{Tabur} {et~al.}(2009){Tabur}, {Bedding}, {Kiss}, {Moon}, {Szeidl},
  \& {Kjeldsen}}]{tabur2009}
{Tabur}, V., {Bedding}, T.~R., {Kiss}, L.~L., {et~al.} 2009, \mnras, 400, 1945

\bibitem[{{Torres}(2010)}]{torres2010}
{Torres}, G. 2010, \aj, 140, 1158

\bibitem[{{Whitelock} {et~al.}(2009){Whitelock}, {Menzies}, {Feast},
  {Matsunaga}, {Tanab{\'e}}, \& {Ita}}]{2009MNRAS.394..795W}
{Whitelock}, P.~A., {Menzies}, J.~W., {Feast}, M.~W., {et~al.} 2009, \mnras,
  394, 795

\end{thebibliography}

%

\newpage

\begin{appendix} 
\section{Photocenter position for the different RHD simulations}

\begin{figure}
\includegraphics[width=0.95\hsize]{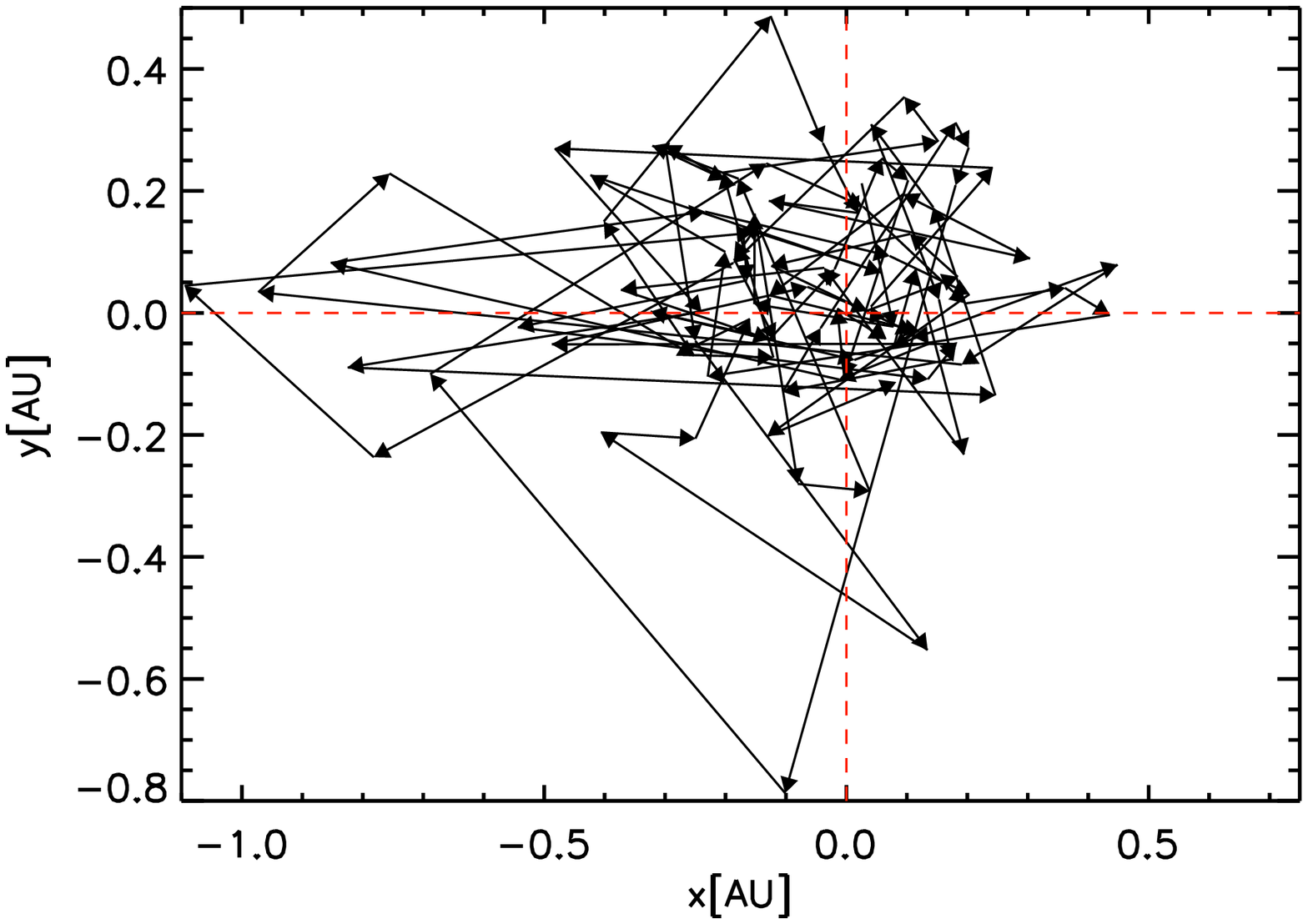}
\caption{Photocenter position computed from RHD simulation st26gm07n001 in Table~\ref{simulations} in the Gaia $G$ band filter. The different snapshots are connected by the line segments, the total time covered is reported in the Table. The dashed lines intersect at the position of the geometrical center of the images.} 
\label{photo1}
\end{figure}

\begin{figure}
\includegraphics[width=0.95\hsize]{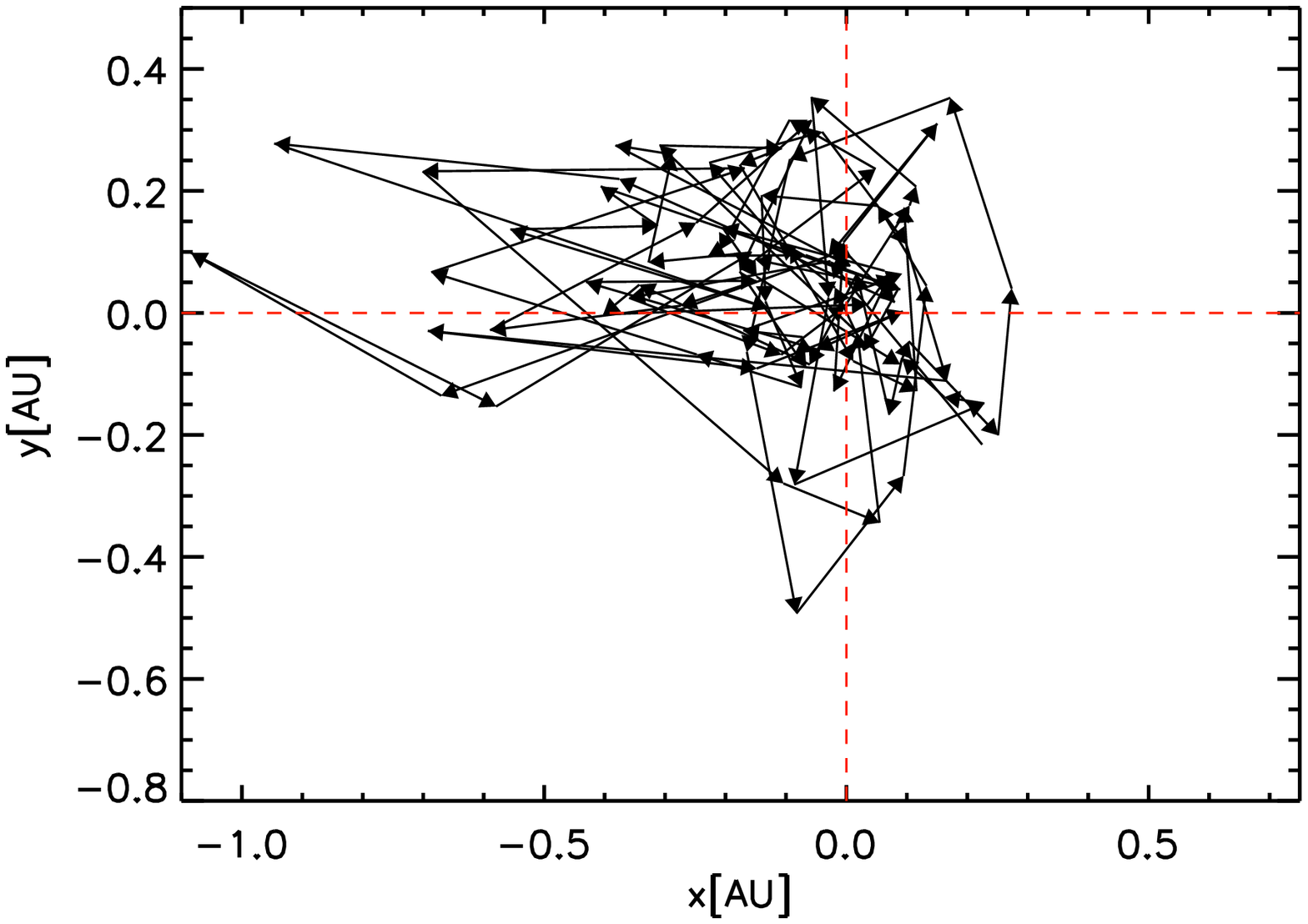}
\caption{Same as in Fig.~\ref{photo1} for RHD simulation st26gm07n002 in Table~\ref{simulations}.} 
\label{photo2}
\end{figure}

\begin{figure}
\includegraphics[width=0.95\hsize]{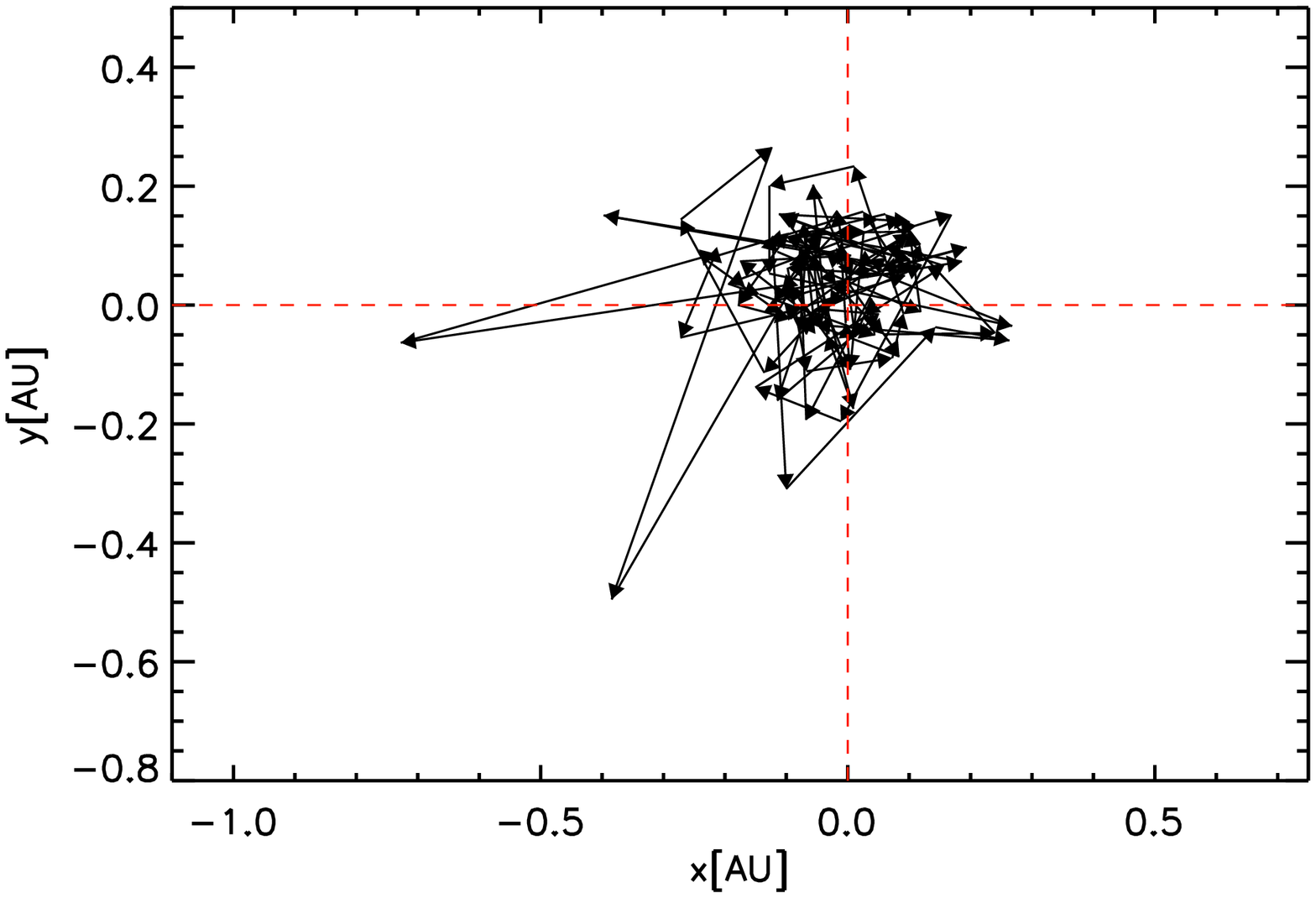}
\caption{Same as in Fig.~\ref{photo1} for RHD simulation st27gm06n001 in Table~\ref{simulations}.} 
\label{photo3}
\end{figure}

\begin{figure}
\includegraphics[width=0.95\hsize]{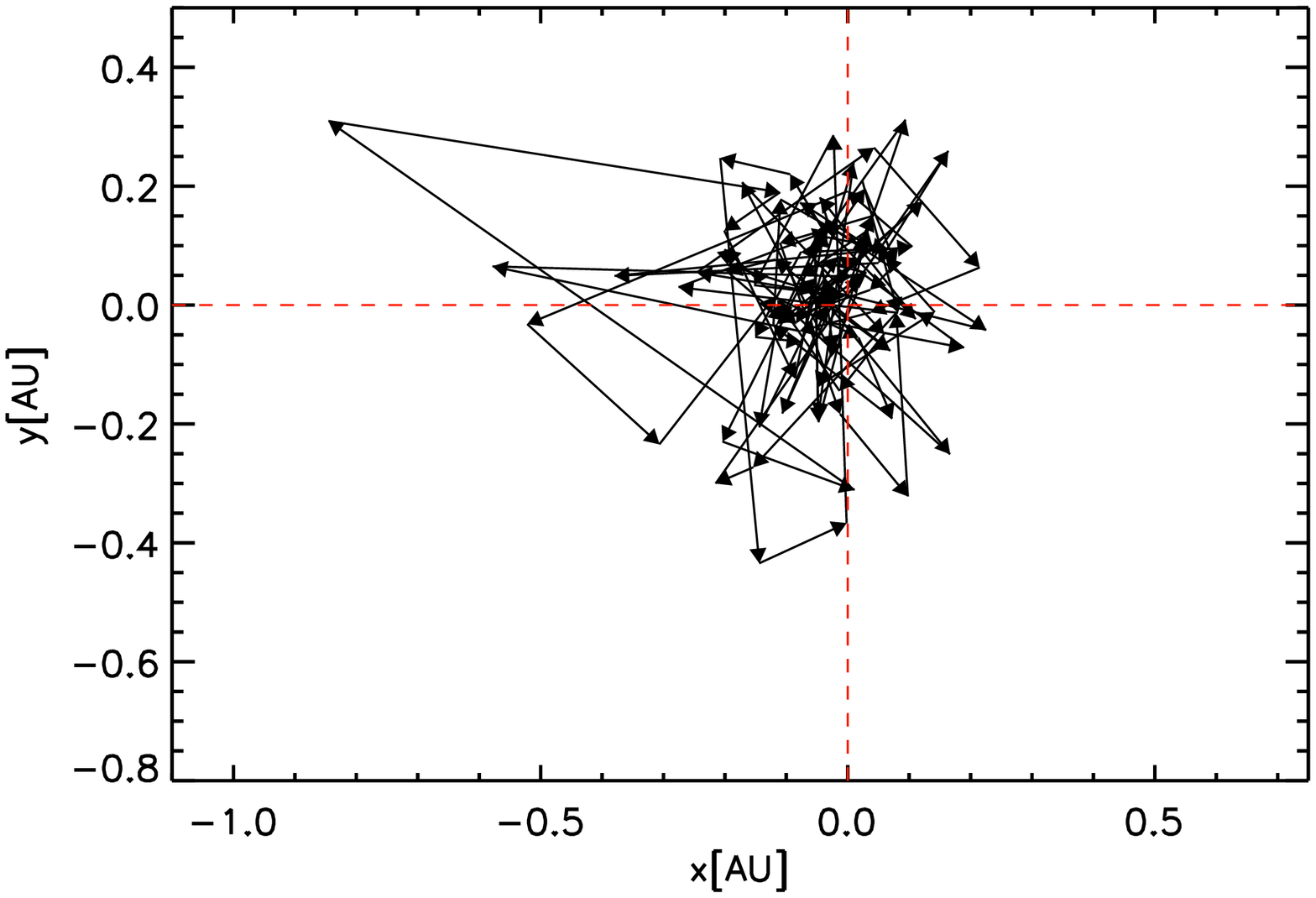}
\caption{Same as in Fig.~\ref{photo1} for RHD simulation st28gm05n001 in Table~\ref{simulations}.} 
\label{photo4}
\end{figure}

\begin{figure}
\includegraphics[width=0.95\hsize]{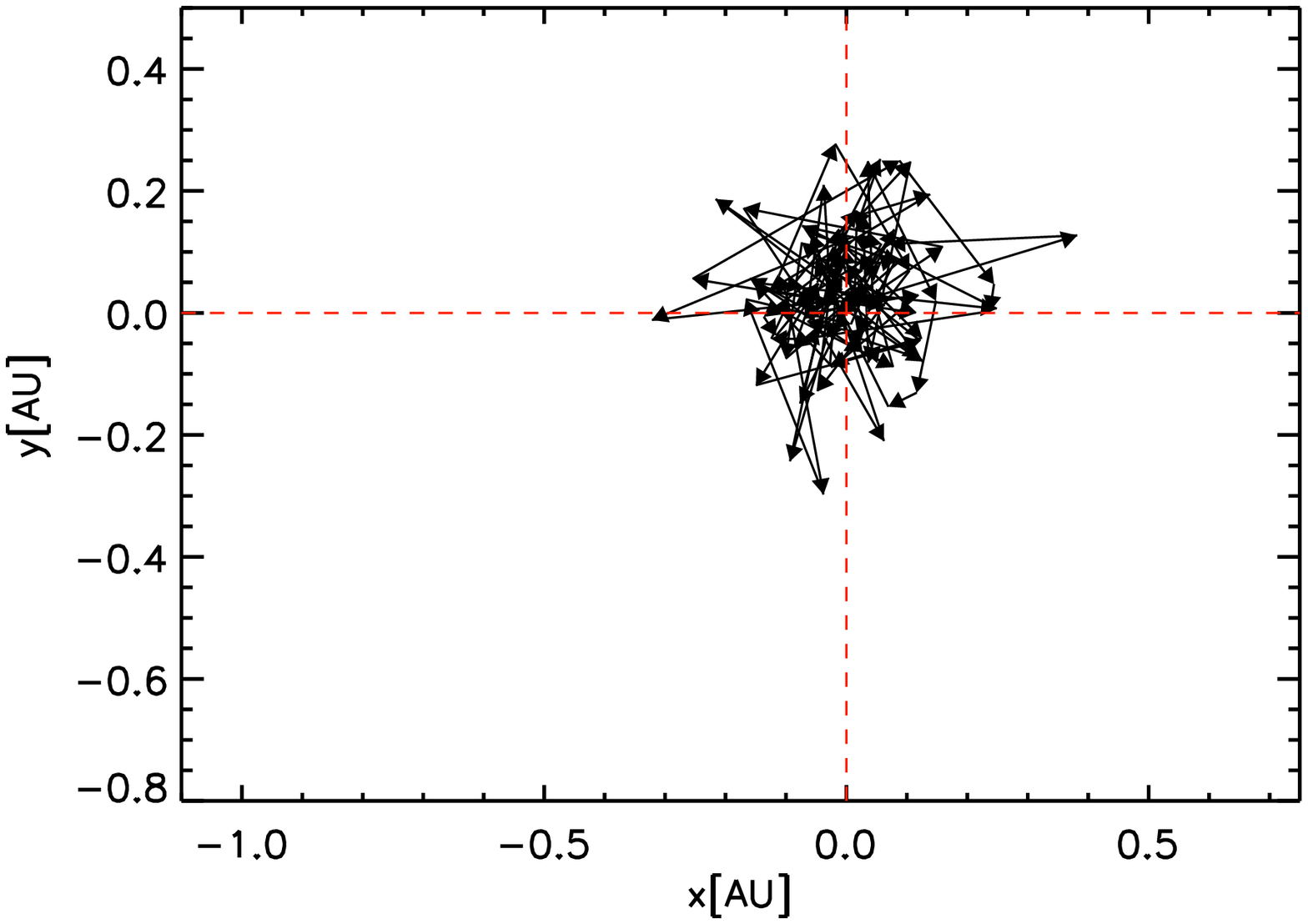}
\caption{Same as in Fig.~\ref{photo1} for RHD simulation st28gm05n002 in Table~\ref{simulations}.} 
\label{photo5}
\end{figure}

\begin{figure}
\includegraphics[width=0.95\hsize]{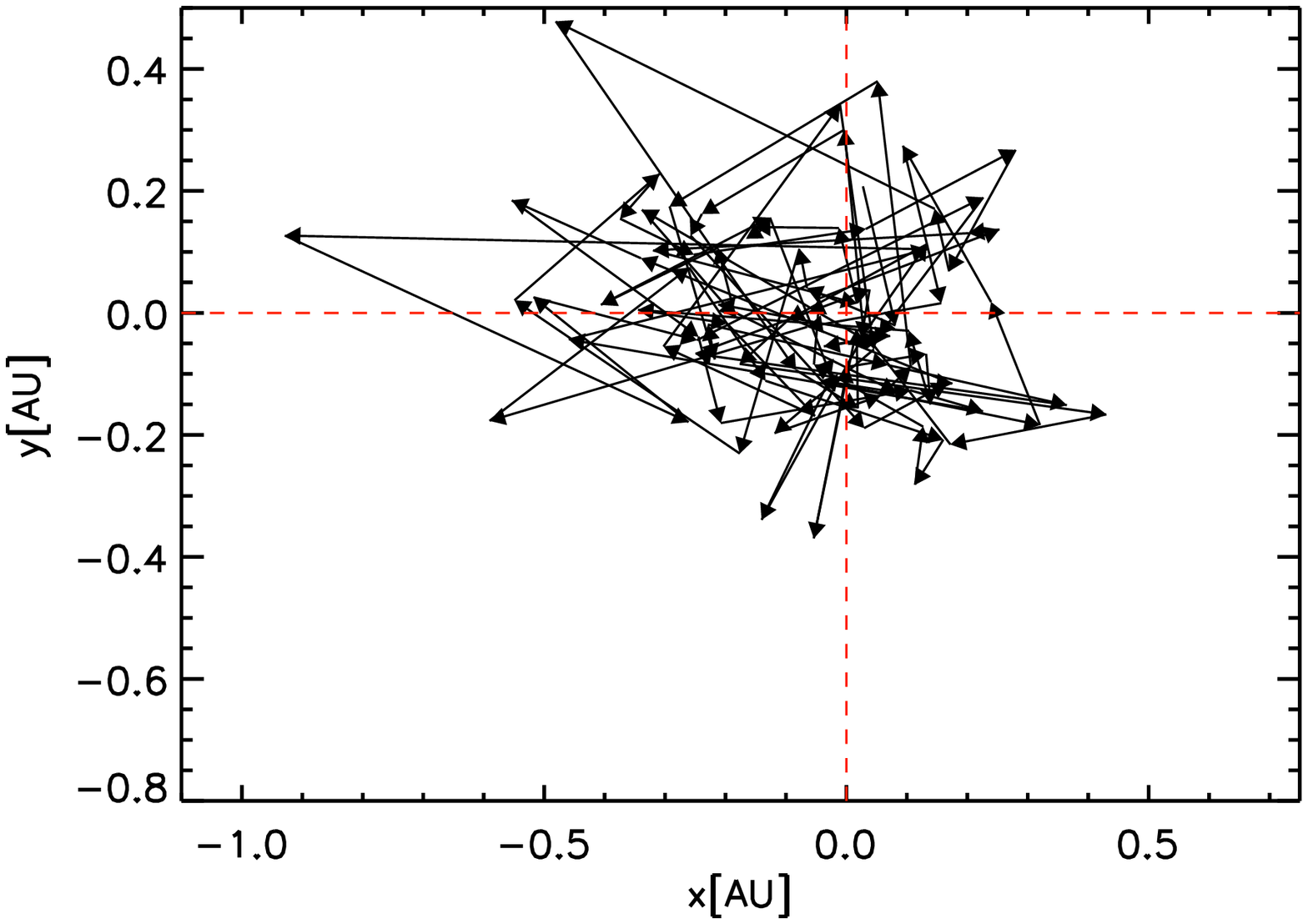}
\caption{Same as in Fig.~\ref{photo1} for RHD simulation st28gm06n26 in Table~\ref{simulations}.} 
\label{photo6}
\end{figure}

\begin{figure}
\includegraphics[width=0.95\hsize]{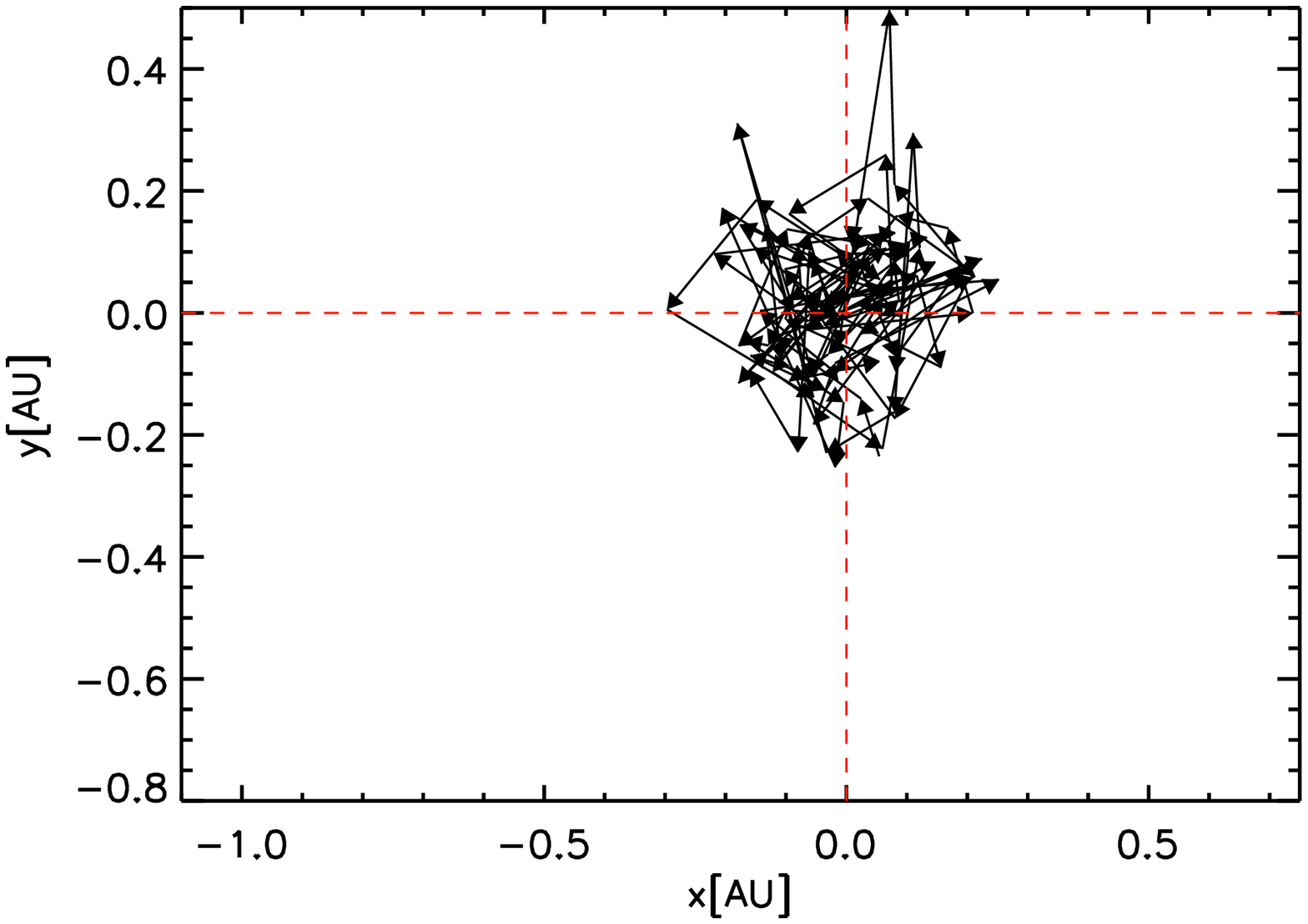}
\caption{Same as in Fig.~\ref{photo1} for RHD simulation st29gm04n001 in Table~\ref{simulations}.} 
\label{photo8}
\end{figure}

\begin{figure}
\includegraphics[width=0.95\hsize]{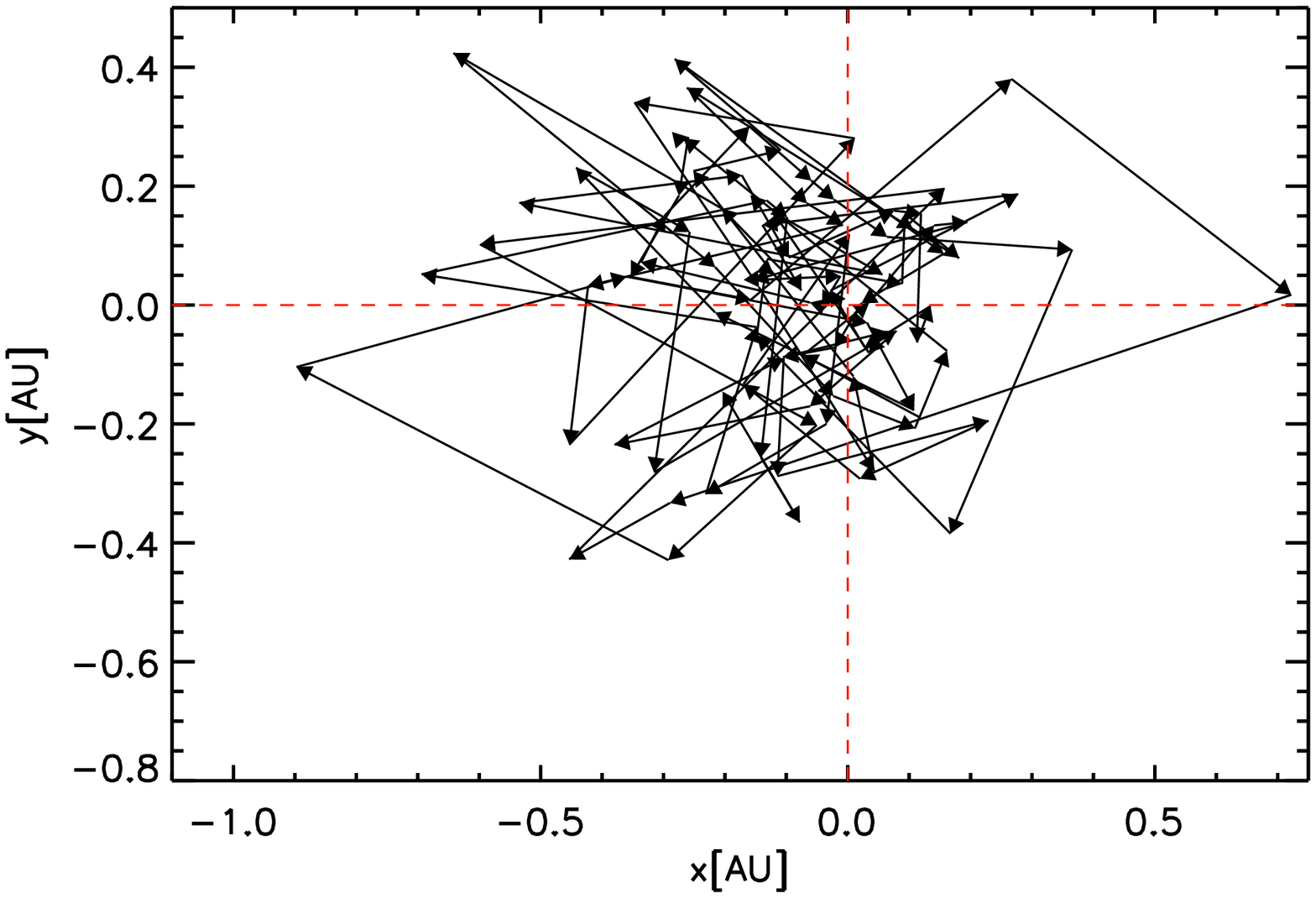}
\caption{Same as in Fig.~\ref{photo1} for RHD simulation st29gm06n001 in Table~\ref{simulations}.} 
\label{photo9}
\end{figure}

\end{appendix}

\end{document}